\documentclass[conference,a4paper]{IEEEtran}
\IEEEoverridecommandlockouts
\usepackage[left=1.1cm,right=1.1cm,top=1.1cm,bottom=1.1cm]{geometry}
\usepackage{mleftright}       
\mleftright                   

\usepackage{graphicx}         
\usepackage{booktabs} 

\usepackage[utf8]{inputenc}
\usepackage{amsmath,amssymb,amsfonts}
\usepackage{mathtools}
\usepackage{amsthm}
\usepackage{algorithmic}
\usepackage{ragged2e}
\usepackage{textcomp}
\usepackage{tikz}
\usepackage{caption}
\usepackage{cuted}
\usepackage{pgfgantt}
\usepackage{pdflscape}
\usepackage{pst-plot}
\usepackage{comment} 
\usepackage{cases}
\usepackage{nccfoots}
\usepackage{lineno}
\usepackage{graphicx}
\usepackage[caption=false,font=footnotesize]{subfig}
\usetikzlibrary{spy}
\usetikzlibrary{positioning,calc}
\usetikzlibrary{decorations.pathmorphing,calc,shapes,shapes.geometric,patterns}
\usetikzlibrary{shapes.multipart}
\usepackage{xfrac}
\usepackage{colortbl}
\usepackage{cancel} 
\usepackage{bbm}
\usetikzlibrary{arrows,positioning,calc,intersections}
\usetikzlibrary{datavisualization.formats.functions}
\def\BibTeX{{\rm B\kern-.05em{\sc i\kern-.025em b}\kern-.08em
    T\kern-.1667em\lower.7ex\hbox{E}\kern-.125emX}}
    
\usepackage{romannum}
\usepackage{pgfplots}
\usepackage{float}
\usepgfplotslibrary{fillbetween}
\usetikzlibrary{arrows, decorations.markings}
\usetikzlibrary{arrows.meta}

\newtheorem{theorem}{Theorem}
\newtheorem*{theorem*}{Theorem}

\newcommand{\up}{u^\prime}
\newcommand{\upp}{u^{\prime\prime}}
\newcommand{\Mp}{M^\prime}
\newcommand{\Mpp}{M^{\prime\prime}}
\newcommand{\dc}{\mathcal{D}}
\newcommand{\dcp}{\mathcal{D}^\prime}
\newcommand{\dcpp}{\mathcal{D}^{\prime\prime}}
\newcommand{\C}{\mathcal{C}}
\newcommand{\Cp}{\mathcal{C}^\prime}
\newcommand{\Cpp}{\mathcal{C}^{\prime\prime}}

\DeclarePairedDelimiterX{\infdivx}[2]{(}{)}{%
  #1\;\delimsize\|\;#2%
}

\pgfplotsset{compat=1.18}
\begin{document}
\title{\huge Security and Privacy: Key Requirements for Molecular Communication in Medicine and Healthcare} 


\author{
\IEEEauthorblockN{Vida Gholamiyan\IEEEauthorrefmark{1}, Yaning Zhao\IEEEauthorrefmark{1}, Wafa Labidi \IEEEauthorrefmark{2}, Holger Boche\IEEEauthorrefmark{2}, and Christian Deppe\IEEEauthorrefmark{1}} 
\IEEEauthorblockA{
\footnotesize\IEEEauthorrefmark{1}Technical University of Braunschweig,\footnotesize\IEEEauthorrefmark{2}Technical University of Munich\\
Email: vidagholamiyan94@gmail.com, yaning.zhao@tu-bs.de, wafa.labidi@tum.de, boche@tum.de, christian.deppe@tu-bs.de}
\thanks{\scriptsize The authors acknowledge the financial support by the Federal Ministry of Education and Research
of Germany (BMBF) in the programme of “Souverän. Digital. Vernetzt.”. Joint project 6G-life, project identification number: 16KISK002 and 16KISK263.
H. Boche and W. Labidi were further supported in part by the BMBF within the national initiative on Post Shannon Communication (NewCom) under
Grant 16KIS1003K and within the national initiative on moleculare communication (IoBNT) under the grant 16KIS1988. C.\ Deppe was further supported in part by the BMBF within the national initiative on Post Shannon Communication (NewCom) under Grant 16KIS1005. C. Deppe, W. Labidi and Y. Zhao were also supported by the DFG within the projects DE1915/2-1 and BO 1734/38-1.}
\vspace{-.8cm}
}

\maketitle

\begin{abstract}
Molecular communication (MC) is an emerging paradigm that enables data transmission through biochemical signals rather than traditional electromagnetic waves. This approach is particularly promising for environments where conventional wireless communication is impractical, such as within the human body. However, security and privacy pose significant challenges that must be addressed to ensure reliable communication. 
Moreover, MC is often event-triggered, making it logical to adopt goal-oriented communication strategies, similar to those used in message identification. This work explores secure identification strategies for MC, with a focus on the information-theoretic security of message identification over Poisson wiretap channels (DT-PWC). 

\end{abstract}
\section{Introduction}
Molecular communication (MC) transmits information using biochemical molecules, making it ideal for environments where electromagnetic waves are ineffective, such as inside living organisms or underwater. A key area of MC research is its integration with the internet of bio-nano-things (IoBNT), enabling nanoscale devices to communicate with larger networks, including future 6G systems \textcolor{black}{\cite{schwenteck20236g}}. Potential applications include targeted drug delivery, bio-sensing, and medical diagnostics. However, MC faces challenges like signal noise, delays, and security risks. Since molecular signals can be intercepted, robust security measures are needed to prevent eavesdropping and data corruption. 

In MC, data transmission is event-triggered, occurring only in response to specific biological or environmental events. This is crucial in intra-body communication, where molecular signals are released based on conditions like biomarker presence, pH changes, or immune responses. Given this nature, goal-oriented strategies ensure efficient and secure communication. One approach is message identification \cite{ahlswede1989identification}, which prioritizes recognizing whether a specific message was sent rather than reconstructing the entire sequence. Unlike traditional methods, it enhances security, energy efficiency, and robustness in noisy, resource-constrained environments, reducing interception risks in adversarial scenarios like Poisson wiretap channels. A key challenge in IoBNT is securing information transmission \cite{haselmayr2019integration}. Without protection, it could be misused to steal health data or create diseases. Combining cybersecurity methods with biological defenses, like the human immune system, can help address this risk. \textcolor{black}{One potential application in MC involves a malicious eavesdropper concealing itself behind a legitimate receiver. This scenario motivates the use of a degraded discrete-time Poisson wiretap channel (DT-PWC).}
\section{Identification and Event-Triggered Molecular Communication}
In the classical transmission scheme proposed by Shannon, the encoder transmits a message over a channel, and at the receiver side, the decoder aims to estimate this message based on the channel observation. However, this is not the case for identification, a new
approach in communications suggested by Ahlswede and Dueck \cite{ahlswede1989identification} in 1989. This new problem with the semantic aspect has enlarged the basis of information theory. In the identification scheme, the encoder sends an identification message (also called identity) over the channel and the decoder is not interested in \emph{what} the received message is, but wants to know \emph{whether} a specific message, in which the receiver is interested, has been sent or not. This concept is critical in applications where quick decision-making is required, such as emergency alerts in nano-medicine. Identification codes grow double exponentially with block length, allowing efficient transmission even in highly noisy environments. Secure identification ensures that an eavesdropper cannot determine whether a message was transmitted to a specific receiver.

\begin{figure}[h]
    \centering
    \begin{minipage}{\linewidth}
        \centering
        \includegraphics[width=0.8\linewidth]{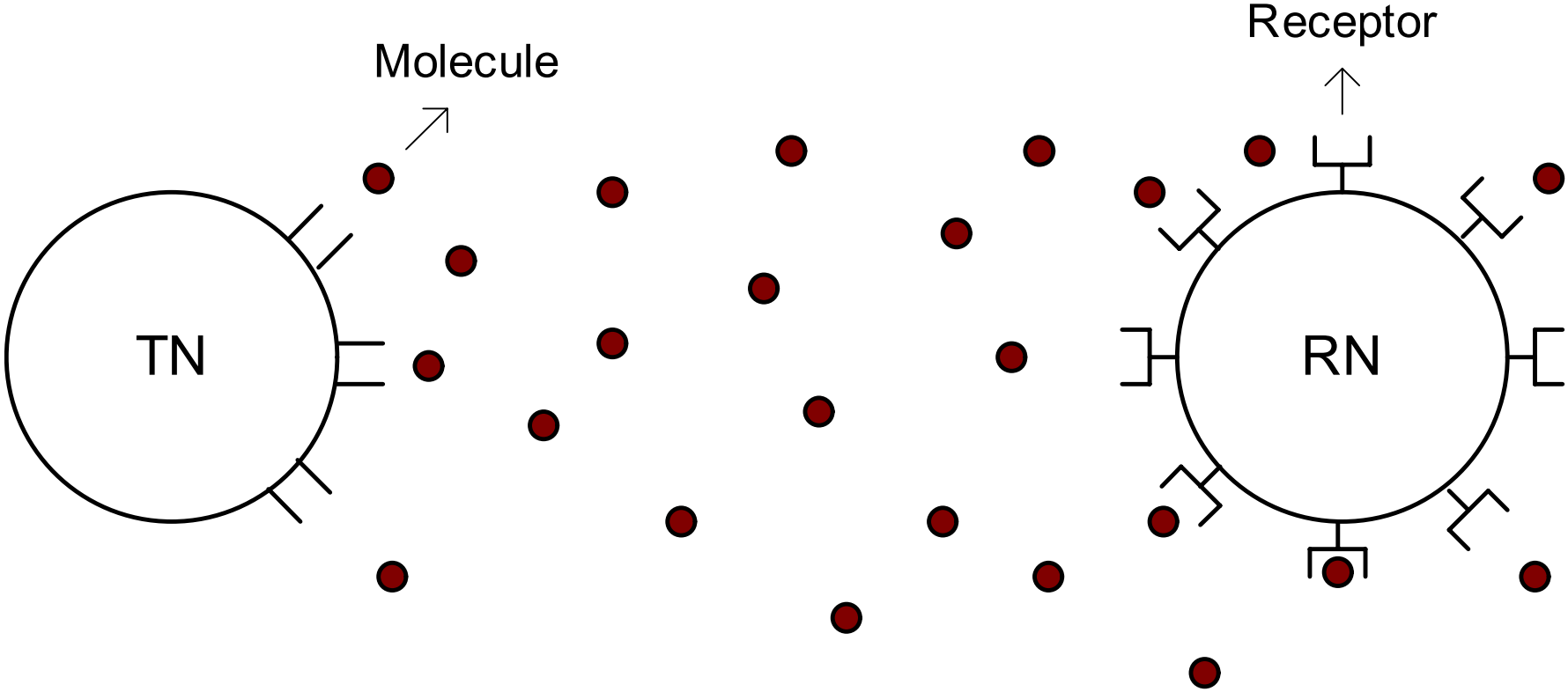}
    \end{minipage}
    
    \begin{minipage}{\linewidth}
        \hspace{-5mm}
        \input{Figures/Poisson}
    \end{minipage}
    \vspace{-0.5em}
    \caption{Event-triggered molecular communication}
    \label{fig:even-triggered MC}
    \vspace{-.5cm}
\end{figure}

MC has shown great potential in identifying and diagnosing abnormalities across various fields, including biotechnology and healthcare, such as in drug delivery, cancer treatment, and health monitoring. Specifically, the use of MC for detecting abnormalities in medical settings has been a key area of research in recent years. Abnormality detection in this context refers to identifying viruses, bacteria, infectious microorganisms, or tumors within the body using either stationary or mobile nanomachines. Once an abnormality is identified, treatment can be initiated through drug delivery systems, targeted therapies, or nanosurgery.
\begin{figure}[h]
    \centering
    \includegraphics[width=0.95\linewidth]{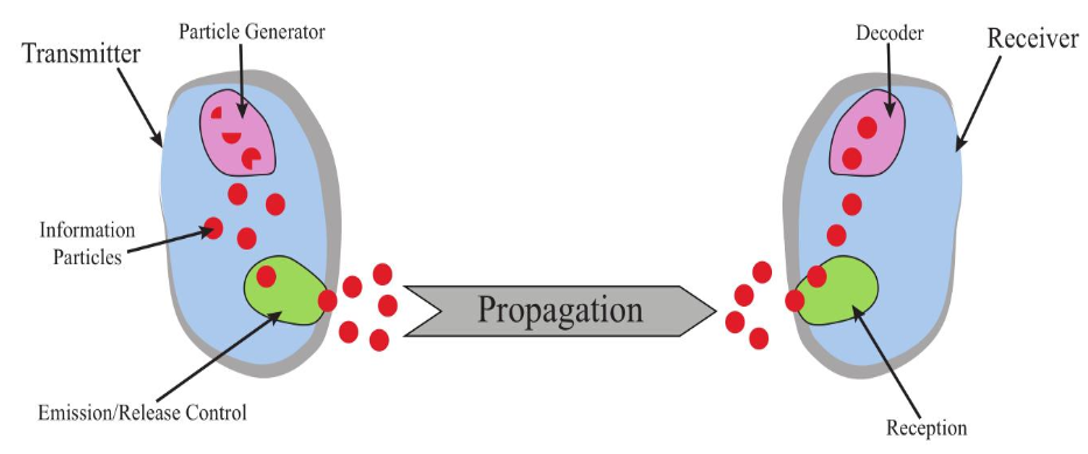}
    \vspace{-0.5em}
    \caption{Molecular communication}
    \label{fig:MC}
\end{figure}
\vspace{-0.5em}
In terms of information theory, diffusion-based MC has been studied to understand its capacity and limitations. The capacity of molecular timing channels in diffusion-based MC has been analyzed, providing both upper and lower bounds for the system's performance.
The stochastic nature of molecular propagation makes event-triggered communication a suitable approach. This model relies on the discrete-time Poisson channel (DTPC) as illustrated in Fig. \ref{fig:even-triggered MC}, where the number of received molecules occurs at random intervals following a Poisson distribution \cite{labidi2023information}. Let a memoryless DTPC $\left(\mathcal{X},\mathcal{Y},W(y|x)\right)$ consisting of input alphabet $\mathcal{X}\subset \mathbb{R}_0^+$, output alphabet $\mathcal{Y}\subset \mathbb{Z}_0^+$ and probability massive function (pmf) $W(y|x)$ on $\mathcal{Y}$, be given. For $n$ channel uses, the transition probability law is given by
\begin{align}
    W^n(y^n|x^n)
    &=\prod_{t=1}^{n}(y_t|x_t)\nonumber\\
    &=\prod_{t=1}^{n}\text{exp}\left(-(x_t+\lambda_0)\right)\frac{(x_t+\lambda_0)^{y_t}}{y_t!},
\end{align}
where $\lambda_0$ is some nonnegative constant, called dark current, representing the non-ideality of the detector. The sequences $x^n=\left(x_1,x_2,\cdots,x_n\right)\in\mathcal{X}^n$ and $y^n=(y_1,y_2,\cdots,y_n)\in\mathcal{Y}^n$ are the channel input and the channel output, respectively. 

This mechanism is particularly effective for low-power, burst-based transmission schemes in biomedical applications. By leveraging event-triggered coding, it is possible to minimize energy consumption while maximizing data security and robustness against interference.

\section{Information-Theoretic Secure Communication}
Security in MC remains a challenge. Research focuses on efficient, low-latency coding and secure molecular transceivers. MC-6G integration demands advances in information theory, nanotechnology, and biochemical cryptography.

Transmission over a degraded discrete-time Poisson wiretap channel (DT-PWC) has been considered in \cite{soltani2021degraded}. The transmission capacity of degraded DT-PWC is given by
\begin{align}
    C_s(W,V)=\max_{P_X} \left(I(X;Y)-I(X;Z)\right).
\end{align}

\begin{figure}[H]
    \centering
    \input{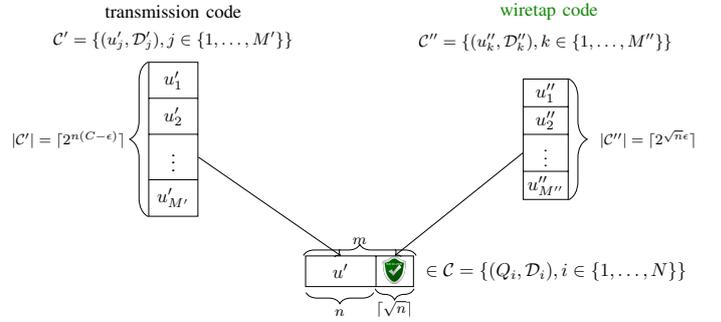}
    \vspace{-0.5em}
    \caption{The construction of the identification wiretap code}
    \label{fig:code}
\end{figure}

\begin{figure}[h]
\hspace{-.8cm}
    \input{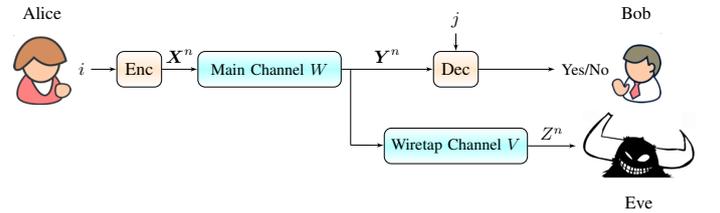} 
    \vspace{-.5cm}
    \caption{Discrete-time Poisson wiretap channel}
    \label{fig:PWC}
    \vspace{-.5cm}
\end{figure}

Secured identification problem has been first considered in \cite{ahlswede1994identification}. Identification over Gaussian wiretap channel has been explored in \cite{labidi2020secure} as follows.
\begin{theorem}
    Let $C(W)$ denote the transmission capacity of the main channel $W$, $C_s(W,V)$ and $C_{s,ID}$ denote the secure transmission capacity and secure identification capacity, respectively. Then
    \begin{align}
        C_{s,ID}(W,V)=
        \left\{
        \begin{array}{cc}
             C(W),& C_s(W,V)>0,  \\
             0,& \text{otherwise}. 
        \end{array}
        \right.
    \end{align}
\end{theorem}
The idea of the optimal secure code scheme for identification achieving the capacity is to concatenate two fundamental codes, as shown in Fig. \ref{fig:code}. The idea of the direct proof is to concatenate two fundamental codes. We consider a transmission code $\Cp$ and a wiretap code $\Cpp$  as depicted in Fig. \ref{fig:code}. 
For the message set $\{1,\ldots,\Mp\}$ one uses $\{1,\ldots,\Mpp\}$ as a suitable indexed set of colorings of the messages with a smaller number of colors. Both of the coloring and color sets are known to the sender and the receiver(s). Every coloring function, denoted by $T_i \colon \Cp \longrightarrow \Cpp$, corresponds to an identification message $i$. The sender chooses a coloring number $j$ randomly from the set $\{1,\ldots,\Mp\}$ and calculates the color of the identification message $i$ under coloring number $j$ using $T_i$, denoted by $T_i(j)$. We send both of $j$ with the code $\Cp$ and $T_i(j)$ with the code $\Cpp$ over the GWC. 
The receiver, interested in the identification message $i$, calculates the color of $j$ under $T_i$
and checks whether it is equal to the received color or not. In the first case, he decides that the identification message is $i$, otherwise he says it was not $i$.
\vspace{-0.5em}
To the best of our knowledge, it is not known whether secure identification can be reached for discrete-time Poisson wiretap channels (DT-PWCs). In our work, we consider the secured identification over DT-PWC as illustrated in Fig. \ref{fig:PWC}. This model consists of of a transmitter (Alice), a legitimate receiver (Bob), and an eavesdropper (Eve). Bob detects these molecules, while Eve, observes a degraded version due to diffusion and environmental noise. The channel input, channel outputs at the legitimate receiver and the eavesdropper are denoted as $X$, $Y$ and $Z$ respectively. The pmfs of the main channel $W(\cdot,\cdot)$ and the main channel and the wiretap channel $V(\cdot,\cdot)$ are given by
\begin{align}
    W(y|x)&= \text{exp}\left(-(x+\lambda_B)\right)\frac{(x+\lambda_B)^{y}}{y!} \\
    V(z|x)&= \text{exp}\left(-(x+\lambda_E)\right)\frac{(x+\lambda_E)^z}{z!},
\end{align}
respectively, where $\lambda_{B}$ and $\lambda_{E}$ are the dark currents at the legitimate user and the eavesdropper's receiver, respectively.

\section{Conclusion}
Ensuring secure communication within MC systems is essential for their widespread adoption. By addressing the challenges of secure identification over DT-PWC, future research can pave the way for robust and reliable MC networks within 6G ecosystems.
\bibliographystyle{IEEEtran}
\bibliography{definitions,references}

\IEEEtriggeratref{4}



\end{document}